\documentclass[prd,nofootinbib,superscriptaddress]{revtex4}
\usepackage{epsfig}
\usepackage{amsmath}
\usepackage{amsfonts}
\usepackage{amssymb}
\usepackage{float}
\usepackage{color}
\usepackage{graphicx}
\usepackage{graphics}
\usepackage{hyperref}
\hypersetup{}
\usepackage{epstopdf}
\usepackage{multirow}

\definecolor{rosso}{cmyk}{0,1,1,0.4}
\definecolor{rossos}{cmyk}{0,1,1,0.55}
\definecolor{rossoc}{cmyk}{0,1,1,0.2}
\definecolor{blu}{cmyk}{1,1,0,0.3}
\definecolor{blus}{cmyk}{1,1,0,0.6}
\definecolor{bluc}{cmyk}{1,1,0,0.1}
\definecolor{verde}{cmyk}{0.92,0,0.59,0.25}
\definecolor{verdec}{cmyk}{0.92,0,0.59,0.15}
\definecolor{verdes}{cmyk}{0.92,0,0.59,0.4}

\begin{document}

\title{\color{blue}Top quark polarization as 
a probe of charged Higgs bosons}

\author{Abdesslam Arhrib}
\email{aarhrib@gmail.com}
\affiliation{D\'{e}partement de Math\'{e}matiques, 
Facult\'{e} des Sciences et Techniques,
Universit\'{e} Abdelmalek Essaadi, B. 416, Tangier, Morocco.}

\author{Adil Jueid}
\email{adil.jueid@sjtu.edu.cn}

\affiliation{INPAC, Shanghai Key Laboratory for Particle Physics and Cosmology,
Department of Physics and Astronomy, Shanghai Jiao Tong University,
Shanghai 200240, China.}

\author{Stefano Moretti}
\email{s.moretti@soton.ac.uk}
\affiliation{School of Physics and Astronomy, University of Southampton,\\
Southampton, SO17 1BJ, United Kingdom.}

\begin{abstract}
We study the production and decay of a heavy charged Higgs boson in the $bg\to tH^-$ and $H^-\to b\bar t$ processes at the Large Hadron Collider (LHC). We show that the chiral structure of the $H^-t\bar b$ vertex entering both stages is sensitive to the underlying Higgs mechanism of Electro-Weak Symmetry Breaking (EWSB) and we specifically demonstrate that one could distinguish between  two  popular realizations of a 2-Higgs Double Model (2HDM) embedding a new $H^\pm$ state, i.e., those with Type-I and -Y Yukawa couplings. The chiral structure of such a vertex, which is different in the two cases, in turn triggers a particular spin state of the top quark which is then transmitted to its decay products. Hence, both inclusive rates and exclusive observables can be used to extract the presence of such a charged Higgs boson state in LHC data.  
\end{abstract}
\maketitle

\section{Introduction}
\label{Introduction}
The top quark, discovered in the mid nineties at the  Tevatron by the D0 \cite{Abachi:1995iq} 
and CDF \cite{Abe:1995hr} collaborations, is the heaviest
elementary particle we know of. As such, it is the most sensitive probe of new physics Beyond the Standard Model (BSM) onsetting at the TeV scale or above it. Further,  due to its large mass, it can only be created as a real object in scattering  
processes taking place in the most recent particle accelerators like the now 
decommissioned Tevatron and the state-of-the-art LHC. 

Thus, in recent years, studies of processes involving top quarks have become very numerous, also thanks to the large amount of data containing them which can be produced and analyzed. In this respect, an intriguing aspect is that the top quark has a very short
lifetime and its decay width  satisfies $\Gamma_t \simeq G_F m_t^3 \gg \Lambda_{QCD}^2/m_t$
implying that it decays immediately, i.e., before hadronization effects  can take
place. Thus, all its  fundamental properties can be probed by studying its decay products (for a 
review, see \cite{Bernreuther:2008ju, Deliot:2014uua, Bernreuther:2015wqa}). 
Consequent measurements are therefore well established and can be used to either test the SM predictions or look for new physics beyond it.

One of the simplest SM extensions, embedding the now established Higgs mechanism of EWSB, is the 2-Higgs Doublet Model (2HDM), which was proposed four decades ago and has eventually found its way as a low energy manifestation of a higher scale fundamental dynamics which could be made manifest in both direct searches for the new Higgs boson states that it predicts, e.g., at the LHC, and indirect tests in flavor physics
(for a theoretical and phenomenological review, see \cite{Branco:2011iw} and  \cite{Akeroyd:2016ymd}, respectively). In this scenario, two complex isodoublets 
are introduced to break the EW symmetry and eventually
produce both fermion and gauge boson masses. The mediators of such mass generation dynamics are five Higgs boson states: two neutral CP-even ones ($h^0$ and $H^0$, with $m_{h^0}<m_{H^0}$), one CP-odd neutral one ($A^0$) and a pair of charged ones $H^\pm$.

The top quark couple to all of these, hence, it is no surprise that top quark processes have been studied extensively within the 2HDM, not least because the structure of the (Yukawa) coupling between such fermion and (pseudo)scalar bosons can reveal the properties of the underlying 2HDM. Such studies concerned not only inclusive 
production cross sections \cite{Stange:1993td, Zhou:1996dx, Hollik:1997hm} but also the possibility to exclusively extract the aforementioned  
couplings \cite{Denner:1992vz, Bernreuther:2008us, Huitu:2010ad,  Arhrib:2016vts}, using the leading SM-like decay 
channel of the top quark as well as its rare and some exotic ones \cite{Eilam:1990zc, Arhrib:2005nx}.
In essence, it was indeed proven that the top quark 
can serve as a discovery probe of the 2HDM and also to discriminate between its different realizations.

It is the purpose of this paper to contribute to this endeavour by exploiting  polarization effects of the top quark, including  $t \bar{t}$ spin correlations (see, e.g., \cite{Mahlon:1995zn, Mahlon:2010gw, Uwer:2004vp} and references 
therein), in studies of 2HDM charged Higgs bosons at the LHC, for the case when  $m_{H^\pm}>m_t$, i.e., of a heavy $H^\pm$ state. By building upon earlier results \cite{Huitu:2010ad}, which postulated a 2HDM Type-II structure (henceforth, 2HDM-II), we complement them by showing that other 2HDM paradigms can similarly be probed, e.g., the 2HDM Type-I (henceforth, 2HDM-I) and 2HDM Type-Y (henceforth, 2HDM-Y or flipped). Furthermore, we will investigate a larger variety of experimental observables that can be used for the above purpose. We will eventually show that the latter can profitably be exploited both as a mean to improve sensitivity of current heavy $H^\pm$ searches and as a post-discovery tool to characterize the extracted signals in terms of the underlying 2HDM.  The underpinning element in this study is the  chiral structure of the $H^-t\bar b$ vertex and how it affects the production and decay of a heavy charged Higgs boson in the $bg\to tH^-$ and $H^-\to b\bar t$ processes at the CERN machine.

The plan of this paper is as follows. In the next section we describe the 2HDM as well as define benchmark points in its parameter space amenable to phenomenological investigation. We then define observables that can be used in  experimental studies. After presenting our numerical results, we conclude.

\section{Model, constraints and benchmark points}
\label{sec:2HDM}
\subsection{The model}
In this section, we discuss briefly the 2HDM and several constraints 
imposed on its parameter space. In this model, two scalar isodoublets
are included to break the EW gauge symmetry and give rise 
to fermion and gauge boson masses. However, due to the presence 
of two Higgs doublets, absence of large tree level 
Flavor Changing Neutral Currents (FCNCs) is not guaranteed unless 
a discrete symmetry, $Z_2$, is imposed \cite{Glashow:1976nt}. Under
this symmetry, scalar fields transform as $H_1 \to H_1$ and $H_2 \to -H_2$,
 hence, four possible combinations of scalar and 
 fermion interactions are possible, which give rise to 
four possible types of 2HDM (see, e.g, \cite{Aoki:2009ha} for more 
details). Hereafter, we define the 2HDM-I the model where  
only $\Phi_2$ couples to all the fermions exactly as in the SM
while the 2HDM Type-II (henceforth, 2HDM-II) is defined such that  $\Phi_2$ couples to up-type quarks and $\Phi_1$ 
to down-type quarks and charged leptons.
In the 2HDM type-X (2HDM-X, called also lepton-specific), the charged leptons couple to $\Phi_1$ 
while all the quarks couple to $\Phi_2$. Finally, the 2HDM-Y case 
is instead built such that $\Phi_2$ couples to up-type quarks and leptons and 
$\Phi_1$ couples to down-type quarks.

The Lagrangian representing the Yukawa interactions is given by:
\begin{eqnarray}
 - {\mathcal{L}_{Yuk}} = \bar{Q}_{L,i} Y_{u,ij} \widetilde{\Phi}_2 u_{R,j} + 
\bar{Q}_{L,i} Y_{d,ij} \Phi_{d} d_{R,j} 
 + \bar{L}_L Y_{l,ij} \Phi_l l_{R,j} + \textrm{h.c.},
 \label{Yukawa}
\end{eqnarray}
where $\Phi_\alpha, \alpha=l, d$ is either $\Phi_1 \textrm{ or } \Phi_2$, 
$Y_{\alpha,ij}$ is a set of $3\times 3$ Yukawa matrices and $i,j=1,2,3$ are the generation indices. 

The most general gauge-invariant, renormalizable and CP-conserving scalar potential, with a softly broken $Z_2$ symmetry, is given by:
\begin{equation} \label{potential}
\begin{split}
V(\Phi_1,\Phi_2) & =  \mu_{11}^2 |\Phi_1|^2 + \mu_{22}^2 |\Phi_2|^2 - \mu_{12}^2 (\Phi_1^\dagger \Phi_2 + \Phi_2^\dagger \Phi_1) +  
               \lambda_1 |\Phi_1|^4 + \lambda_2 |\Phi_2|^4 + \lambda_3 |\Phi_1|^2|\Phi_2|^2  \\ 
 & + \lambda_4 |\Phi_1^\dagger \Phi_2|^2 + \frac{\lambda_5}{2} [(\Phi_1^\dagger \Phi_2)^2 + \text{h.c.} ].
\end{split}
\end{equation}
The parameters $\mu_{11,22}^2, \lambda_{i},~(i=1\ldots4)$ 
are real valued while the parameters $\mu_{12}^2$ and $\lambda_5$
could be complex valued in the case of CP-violation.
The two Higgs doublets $\Phi_1$ and $\Phi_2$ are parametrized  as follows:
\begin{eqnarray}
\Phi_i = \left (\begin{array}{c}
\phi_i^+ \\
v_i + \frac{1}{\sqrt{2}}(h_i + i \omega_i) \\
\end{array} \right),
\qquad  i = 1,2,
\label{Doublet}
\end{eqnarray}
where $v_1$ and $v_2$ are the Vacuum Expectation Values (VEVs)
of the two Higgs doublets. After EWSB, three degrees of freedom are absorbed by the longitudinal gauge bosons $W^\pm$, $Z^0$ and we are left
 with the aforementioned five (pseudo)scalar degrees of freedom as real massive particles: $h^0$, $H^0$,  $A^0$ and $H^\pm$.
These mass eigenstates are obtained through the following relations:
\begin{eqnarray}
 \left ( \begin{array}{c}
 h_1 \\
 h_2 \\
 \end{array} \right) = R(\alpha)
 \left ( \begin{array}{c}
 H^0 \\
 h^0 \\
 \end{array} \right), \quad 
 \left ( \begin{array}{c}
 \phi_1^\pm \\
 \phi_2^\pm \\
 \end{array} \right) = R(\beta)
 \left ( \begin{array}{c}
 G^\pm \\
 H^\pm \\
 \end{array} \right), \quad 
 \centering \left ( \begin{array}{c}
 \omega_1 \\
 \omega_2 \\
 \end{array} \right) = R(\beta)
 \left ( \begin{array}{c}
 G^0 \\
 A^0 \\
 \end{array} \right), 
\end{eqnarray}
where $R(\theta)$ is an orthogonal matrix defined as
\begin{eqnarray}
R(\theta) = \left( \begin{array}{c r}
                             \cos \theta & - \sin \theta \\
                             \sin \theta & \cos \theta \\
                            \end{array} \right),
\end{eqnarray}
with $\beta$  defined via $\tan \beta = v_2 /v_1$ and $\alpha$  the mixing angle between the CP-even interaction states $h_i$ ($i=1,2$). The  $\omega_i$ and $\phi_{i}^\pm$ ($i=1,2$) interaction states give in turn rise to the CP-odd Higgs, charged Higgs and Goldstone modes. 

In terms of Higgs mass eigenstates, from the Yukawa Lagrangian in eq.(\ref{Yukawa}) we get
\begin{eqnarray}
 - {\mathcal{L}}_{Yuk} = \sum_{\psi=u,d,l} \left(\frac{m_\psi}{v} \kappa_\psi^h \bar{\psi} \psi h^0 + 
 \frac{m_\psi}{v}\kappa_\psi^H \bar{\psi} \psi H^0 
 - i \frac{m_\psi}{v} \kappa_\psi^A \bar{\psi} \gamma_5 \psi A^0 \right) + \nonumber \\
 \left(\frac{V_{ud}}{\sqrt{2} v} \bar{u} (m_u \kappa_u^A P_L +
 m_d \kappa_d^A P_R) d H^+ + \frac{ m_l \kappa_l^A}{\sqrt{2} v} \bar{\nu}_L l_R H^+ + {\rm h.c.} \right),
 \label{Yukawa-1}
\end{eqnarray}
where the $\kappa_i$'s are the Yukawa couplings in the 2HDM, $V_{ud}$ is a Cabibbo-Kobayashi-Maskawa (CKM) matrix element while  $ P_{L/R}=(1 \mp \gamma_5)/2$ are the Left (L) and Right (R) chiral projectors. We give in Tab. \ref{Yukawa-2} the values of the couplings in the
four types of Yukawa interactions of the 2HDM.
\begin{table}
 \begin{center}
  \begin{tabular}{l|l|l|l|l|l|l|l|l|l}
   \hline \hline
    & $\kappa_u^h$ & $\kappa_d^h$ & $\kappa_l^h$ & $\kappa_u^H$ & $\kappa_d^H$ & $\kappa_l^H$ & $\kappa_u^A$ & $\kappa_d^A$ & $\kappa_l^A$ \\ \hline
    Type-I & $c_\alpha/s_\beta$ & $c_\alpha/s_\beta$& $c_\alpha/s_\beta$ & $s_\alpha/s_\beta$ & $s_\alpha/s_\beta$ & $s_\alpha/s_\beta$ & $\cot\beta$ & 
    $-\cot\beta$ & $-\cot\beta$ \\ \hline
    Type-II & $c_\alpha/s_\beta$ & $-s_\alpha/c_\beta$& $-s_\alpha/c_\beta$ & $s_\alpha/s_\beta$ & $c_\alpha/c_\beta$ & $c_\alpha/c_\beta$ & $\cot\beta$ & 
    $\tan\beta$ & $\tan\beta$ \\ \hline 
    Type-X & $c_\alpha/s_\beta$ & $c_\alpha/s_\beta$& $-s_\alpha/c_\beta$ & $s_\alpha/s_\beta$ & $s_\alpha/s_\beta$ & $c_\alpha/c_\beta$ & $\cot\beta$ & 
    $-\cot\beta$ & $\tan\beta$ \\ \hline
    Type-Y & $c_\alpha/s_\beta$ & $-s_\alpha/c_\beta$& $c_\alpha/s_\beta$ & $s_\alpha/s_\beta$ & $c_\alpha/c_\beta$ & $s_\alpha/s_\beta$ & $\cot\beta$ & 
    $\tan\beta$ & $-\cot\beta$ \\ \hline \hline
    \end{tabular}
 \end{center}
 \caption{Yukawa couplings in terms of mixing angles in the 2HDM-I, -II, -X
   and  -Y.}
 \label{Yukawa-2}
\end{table}
From this, it is clear that, as far as only the charged Higgs couplings to quarks are concerned, there is no difference between 2HDM-I and 2HDM-X or between 2HDM-II and 2HDM-Y.

Finally note that the Higgs sector of the 2HDM has 7 independent parameters, 
which can be taken as  $\tan \beta$, $\sin(\beta-\alpha)$, 
$\mu_{12}^2$ and the four physical Higgs masses. 

\subsection{Constraints and benchmark points} 
The experimental data collected so far at the LHC regarding the 125 GeV Higgs boson discovered in 2012 seem to indicate that the couplings of such a particle to SM objects 
 are to a large extent similar to those in the SM, hence such a Higgs state
 is SM-like. Therefore, any physics Beyond the SM (BSM) must contain a SM-like Higgs boson.  This puts severe constraints on the parameter space of the various 2HDM types  
 and mostly (though not always) 
 pushes the generic model close to its decoupling limit where $h^0$ 
 mimics the SM-like Higgs boson and the other states are rather heavy. The decoupling limit is characterized with 
 $\sin(\beta-\alpha)\approx 1$ ($\beta-\alpha\approx \pi/2$) and $m_{H^0}, m_{A^0}, m_{H\pm}\gg m_{Z^0}, m_{W^\pm}$ \cite{Gunion:2002zf}.
 
 In our analysis, we indeed require that the lightest CP-even scalar is the observed Higgs-like boson with $m_{h^0}=125$ GeV and assume that $\beta-\alpha\approx \pi/2$, 
 which automatically implies that $h^0$ couplings to SM particles are SM-like. The other (pseudo)scalars are chosen nearly degenerate in order to avoid constraints from EW precision measurements. At any rate, all the parameters of the model are subject to several theoretical and experimental constraints. 
The benchmark points that we will use in our study are selected in such a way that 
they satisfy all of these.  As we are concerned with a charged Higgs state, we start by dwelling at some length on the most relevant constraints on its mass and couplings.

The charged Higgs mass is subject to a number of constraints from several $B$-physics observables as well as from direct experimental searches at the LHC (and previous colliders). 
The most stringent flavor bound comes from the $B$ meson Branching Ratios (BRs), chiefly,  
BR$(B\rightarrow X_s \gamma)$ \cite{Misiak:2017bgg}.
For instance, in the 2HDM-II, the current measurement of  BR$(B \to X_s\gamma)$ 
forces the charged Higgs boson mass to be larger than about 580 GeV \cite{Misiak:2017bgg}
while in the 2HDM-I one can still obtain a $H^\pm$ with a 
mass as low as  $100-200$ GeV  provided that $\tan \beta \geq 2$.
Note that, in type-II and -Y, if we allow $b\to s \gamma $ measurement at the 3$\sigma$ level
then it would imply a reduction on the charged Higgs bound from 580 GeV down to 440 GeV \cite{Arhrib:2017veb}.


From the direct search  side, the combined void searches from all four LEP collaborations imply the lower limit $m_{H^\pm} > 78.6$ GeV at 95\% Confidence Level (CL), which applies to all models in which BR$(H^\pm \to \tau \nu)+{\rm  BR}(H^\pm \to cs)=1$ \cite{Schael:2006cr}. Searches for a light charged Higgs boson at the 
Tevatron and LHC have instead been performed from top quark decays: $t\to b H^\pm$ followed by $H^\pm \to \tau \nu$ or $H^\pm \to cs$.  A search for heavy charged Higgs has been also 
 conducted  through $pp\to t H^\pm +X$ with $H^\pm \to \tau \nu$. 
For light $H^\pm$ states, the ATLAS and CMS experiments have already drawn an exclusion on 
BR$(t\to b H^\pm)\times {\rm BR}(H^\pm \to \tau \nu)$ \cite{Akeroyd:2016ymd,Aad:2014kga,Khachatryan:2015uua}.
In the context of some specific MSSM scenarios, these results exclude nearly all $\tan\beta\geq 1$ and  $H^\pm$ masses in the range 80 -- 160 GeV. For a heavy $H^\pm$, they exclude a region of parameter space with high $\tan\beta$ for $H^\pm $ masses between 200 GeV and 250 GeV \cite{Aad:2014kga,Khachatryan:2015uua}.

However, one need to keep in mind the following.
\begin{itemize}
\item In the 2HDM framework, the above limits, which are based on 
fermionic decays of the charged Higgs state, can be weakened if any of the bosonic 
decays $H^\pm \to W^\pm h^0$ and/or $H^\pm \to W^\pm A^0$ are open \cite{Arhrib:2016wpw}.

\item Limits on charged Higgs bosons decaying to $\tau\nu$ final states valid for the 2HDM-II
might be invalidated in the framework of the 2HDM-Y where the $H^\pm\to \tau \nu$ decay rate behaves like $1/\tan\beta$ for large $\tan\beta$.  
This also applies to heavy $H^0/A^0$ searches in the $\tau^+\tau^-$ channel 
because also the $H^0\tau^+\tau^-$ and $A^0\tau^+\tau^-$ couplings are proportional to $1/\tan\beta$ 
for large $\tan\beta$ \cite{Arhrib:2015gra}. A detail analysis of the status of charged Higgs 
in the  four types 2HDM can be found in \cite{Arbey:2017gmh}.
\end{itemize}

As intimated, in our scenario, we will assume that $H^0$, $A^0$ and $H^\pm$ states are nearly  degenerate in mass 
so that $H^\pm \to W^\pm H^0/W^\pm A^0$ decays are closed. Furthermore, since the coupling 
$H^\pm W^\pm h^0$ is proportional to $\cos(\beta-\alpha)$, which is very small in our 2HDM scenarios,
the decay  $H^\pm \to W^\pm h^0$ will be suppressed. Therefore, the decays $H^\pm\to tb$
 and $H^\pm\to \tau \nu$ proceeds with a almost $100\%$ cumulative BR.

We have used the public code \textsc{2HDMC} \cite{Eriksson:2009ws} to make a scan on 
$\mu_{12}^2$, $m_{H^\pm}$, $\tan\beta$, $\sin(\beta-\alpha)$ and the Higgs masses (with the boundary condition that $m_{h^0}=125$ GeV). 
The code also allows for the calculation of all relevant charged Higgs BRs and 
checks several theoretical constraints such as boundedness from below of the scalar potential, tree level perturbative unitarity as well as the EW  precision observables $S$ and $T$. Regarding the latter, the aforementioned 
requirement of near mass degeneracy amongst $H^0$, $A^0$ and $H^\pm$ (specifically, in our scan, that 
$m_{H^0}=m_{H^\pm}+4$ GeV = $m_{A^0}+2$ GeV), implemented in order to suppress 
$H^\pm \to W^\pm H^0/W^\pm A^0$ decays, turns out to be useful to also evade $T$ 
 parameter limits, which can easily be achieved whenever $m_{H^\pm}\approx m_{A^0}$.

The compatibility of the other $h^0$ properties 
with those of the observed SM-like Higgs boson has been checked using \textsc{HiggsSignals} \cite{Bechtle:2013xfa}  through a $\chi^2$ minimization of the Higgs boson signal strengths. Constraints from void searches for extra Higgs bosons at LEP,  Tevatron and the LHC have been applied using \textsc{HiggsBounds} \cite{Bechtle:2013wla}. Herein, a parameter 
  point is excluded at $95\%$ CL if the ratio $r_{95\%}$ between model predictions and data defined by
\begin{eqnarray}
 r_{95\%} = \frac{S_\text{model}}{S_\text{observed}^{95\%}}
\end{eqnarray} 
 is larger than $1$.

\begin{figure}[!t]
\centering
\includegraphics[width=0.48\linewidth]{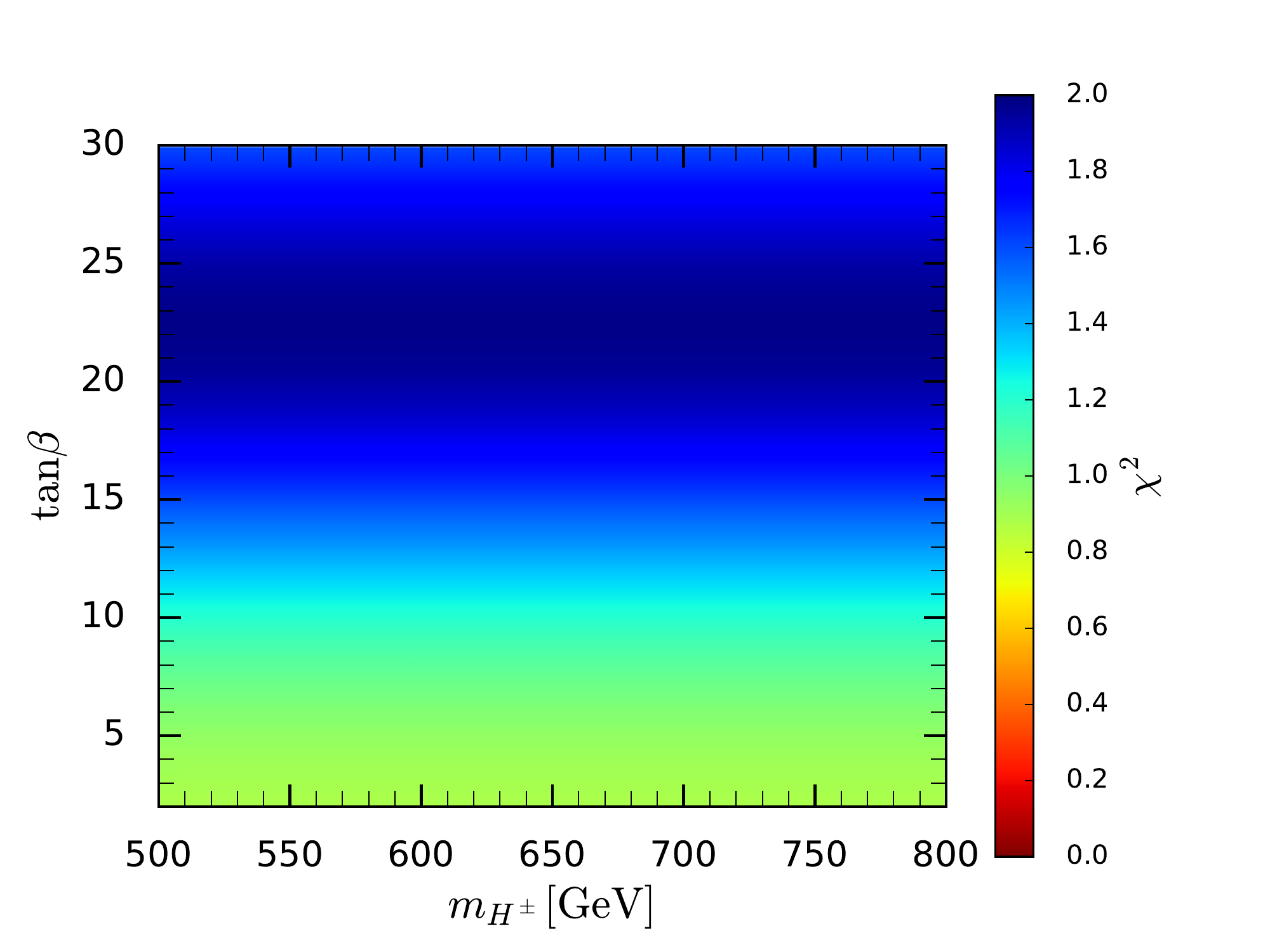}
\hfill
\includegraphics[width=0.48\linewidth]{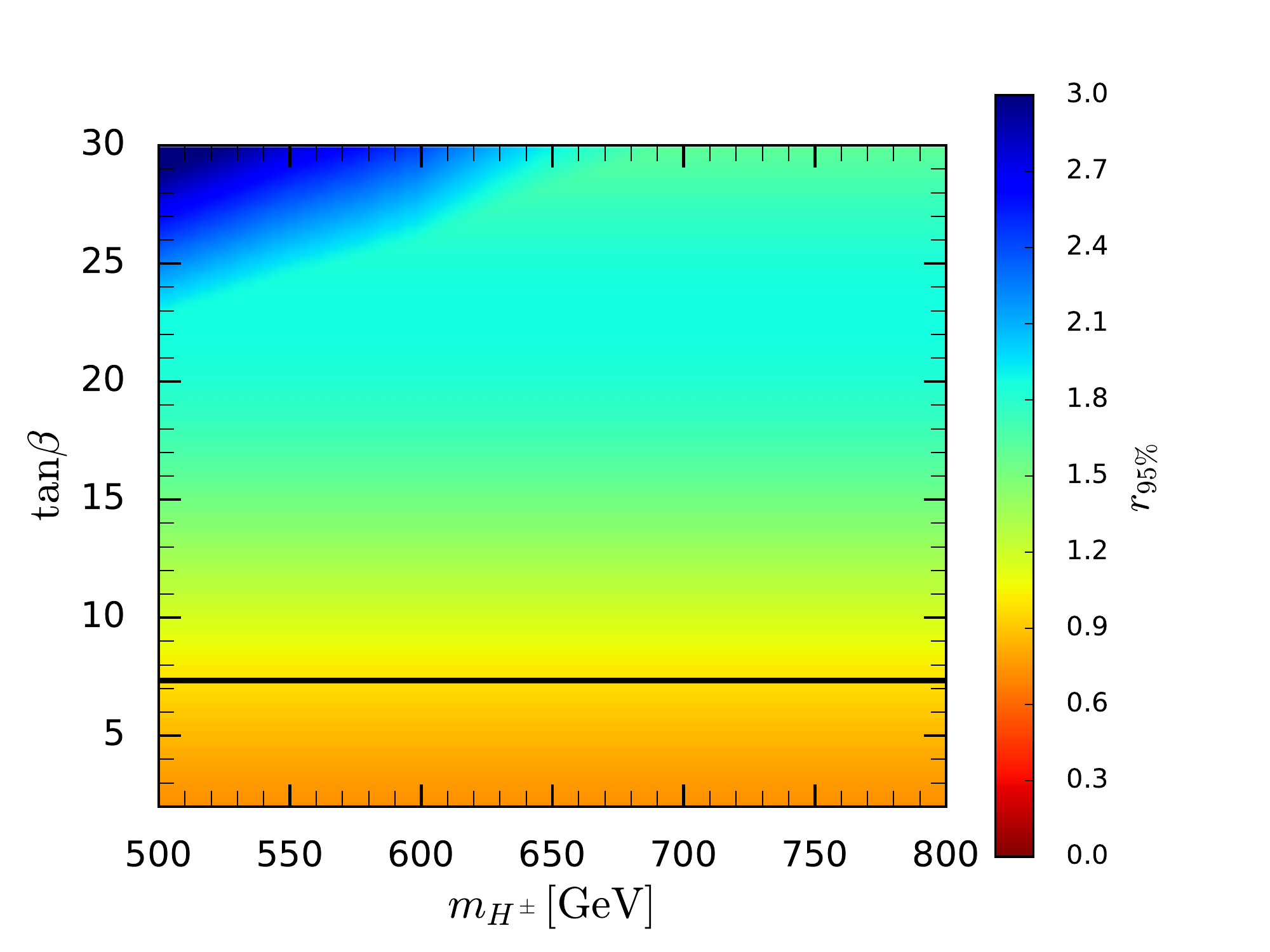}
\caption{\emph{Left}: Scatter plot of $\tan\beta$ and $m_H^\pm$ showing the 
constraints from Higgs boson signal strengths and mass measurements using \textsc{HiggsSignals}. \emph{Right}: The effect of the void direct searches for 
 additional (pseudo)scalars on $\tan\beta$ and $m_{H^\pm}$ using \textsc{HiggsBounds}. The $\chi^2$ (from \textsc{HiggsSignals})
 and 
 $r_{95\%}$ (from \textsc{HiggsBounds}) values  are
  shown as a color map. The horizontal black line on the right 
 shows $r_{95\%}=1$ above which the parameter point is excluded at $95\%$ CL. We illustrate here the case of the 2HDM-II.}
\label{LHC-Higgs-type-II}
\end{figure}

In Fig. \ref{LHC-Higgs-type-II}, we show a scatter plot on  the 
($\tan\beta, m_{H^\pm}$) plane illustrating the impact of experimental constraints on, e.g., the  2HDM-II. In the left panel  we show 
the $\chi^2$ behavior while in the right panel
we show the $r_{95\%}$ one. 
From the first plot, we can see clearly that the SM-like Higgs boson  measurements are consistent with 2HDM-II at the $2\sigma$ level whereas, from the second plot, we can deduce that the absence of discovery of additional (pseudo)scalars exclude $\tan\beta > 8$ for all the charged Higgs masses at $95\%$ CL. Here, the strongest constraint comes from $gg\to A^0/H^0 \to \tau^+\tau^-$ since this process in the 2HDM-II 
is $\tan\beta$ enhanced. As for the 2HDM-I, the effect of the Higgs signal strength measurements is similar to that of the 2HDM-II. However, the effect of void searches  for additional 
Higgs states does  not exclude any points in this case, since the $A^0/H^0$ to $\tau^-\tau^+$ decay rates  are 
 proportional to $1/\tan\beta$, as remarked upon already. Also in the case of the 2HDM-Y such a  $1/\tan\beta$ dependence enable large $\tan\beta$ values, up to 50 or so, so long that $H^\pm$ is heavy enough. 
 Therefore, in the mass region of interest, $m_{H^\pm}>m_t+m_b\approx 180$ GeV, where $H^\pm\to tb$ decays are dominant, both the 2HDM-I and -Y offer considerable regions of available parameter space with 
 the notable difference (which we will exploit in the remainder) that the former can attain small $\tan\beta$ values that are instead precluded to the latter.

\subsection{$t H^\pm$ production in the 2HDM}     
\begin{figure}[!t]
 \centering
 \includegraphics[width=0.48\linewidth]{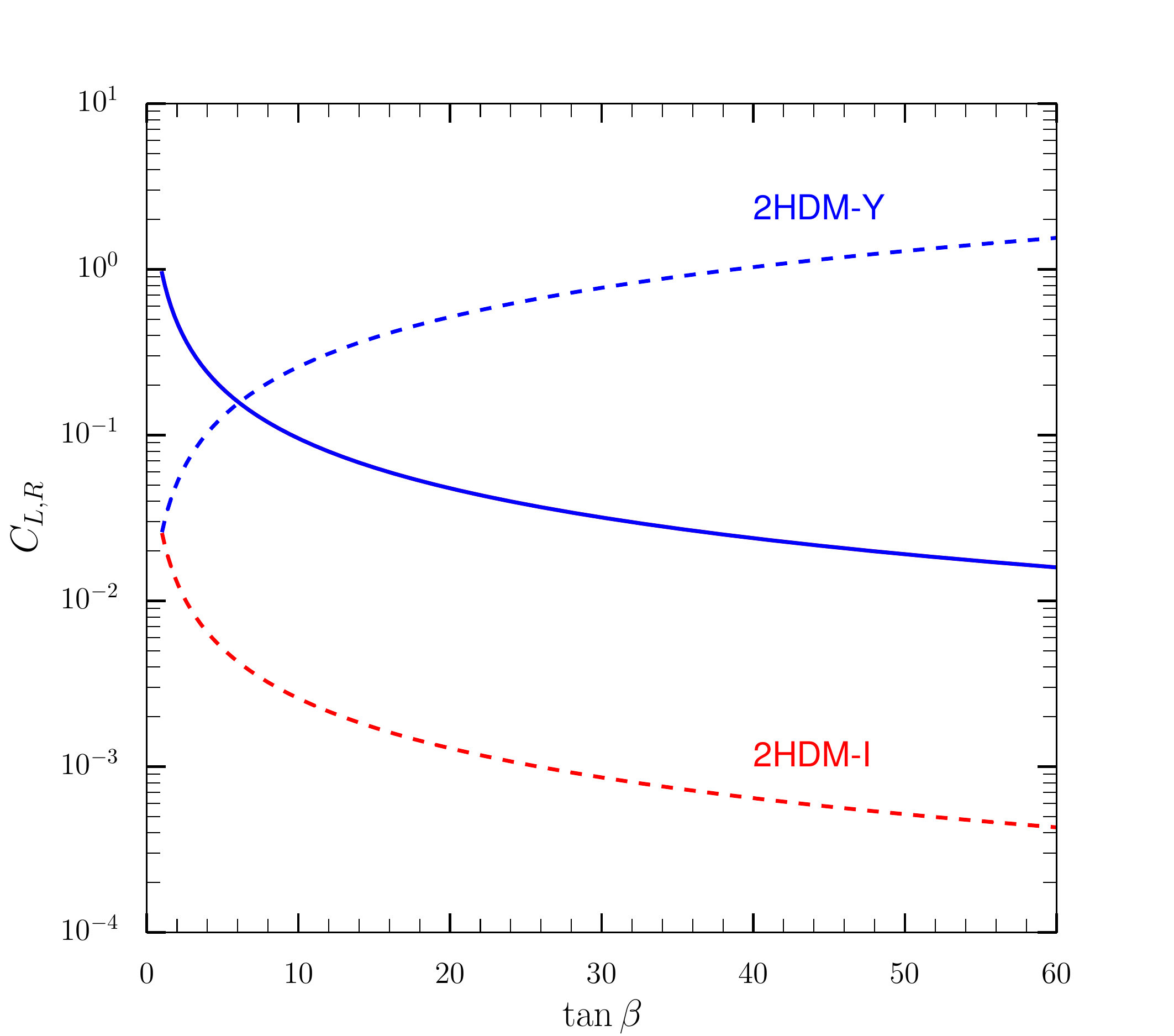}
 \hfill
 \includegraphics[width=0.48\linewidth]{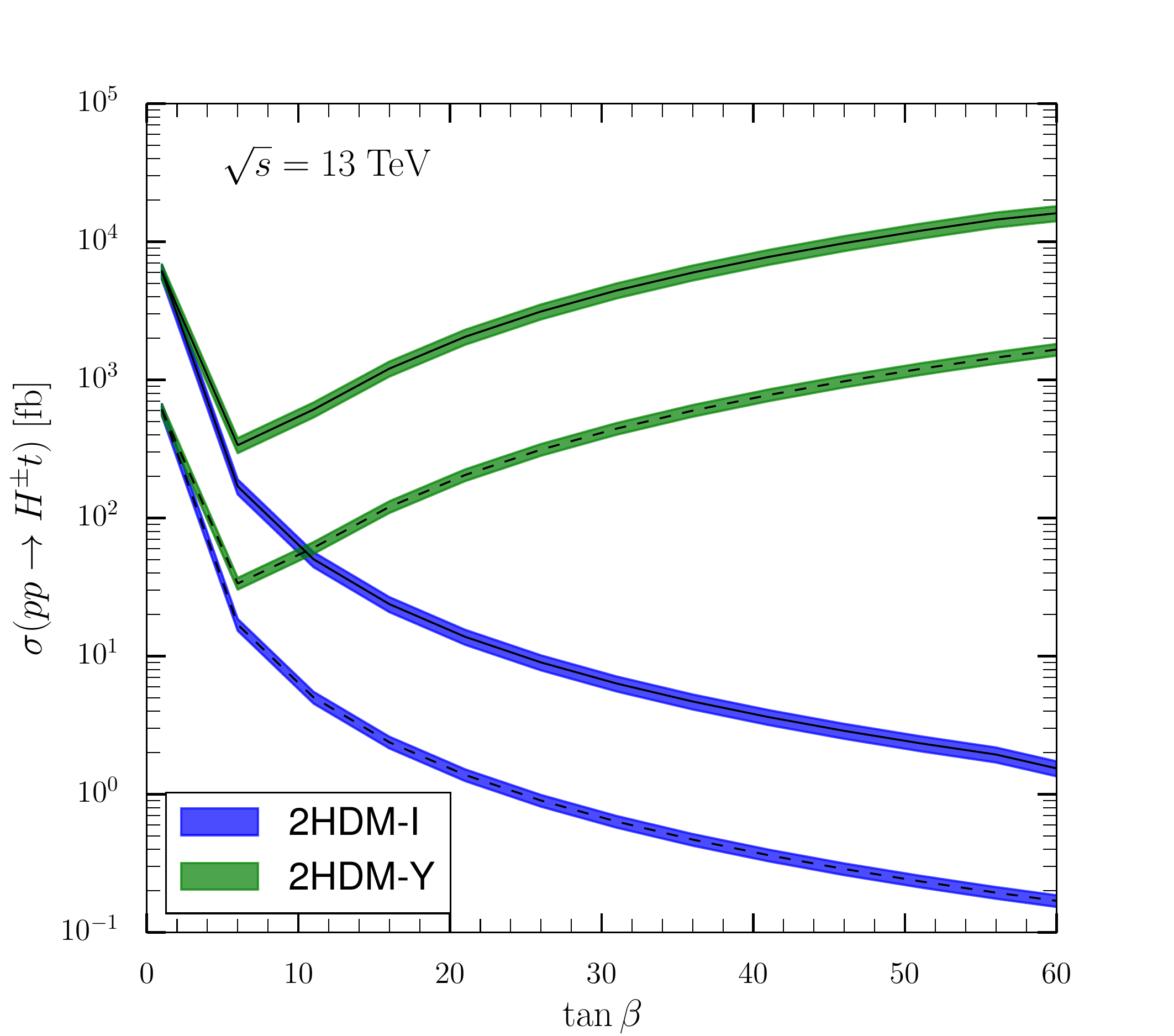}
 \caption{
 \emph{Left}: The chiral form factors of the $H^+\bar tb$ vertex as a function of $\tan\beta$ for the 2HDM-I (red) and 2HDM-Y (blue), where the overlapping solid(separate dashed) lines show the L(R)-handed components. 
 \emph{Right}: The $H^\pm $ production cross section as a function of $\tan\beta$ in the 2HDM-I (blue) and 2HDM-II (green), where the 
 solid(dashed) lines show the cross sections for $m_{H^\pm} = 200~  \textrm{GeV}$($m_{H^\pm} = 500 \textrm{ GeV}$). Here, the error band is a result of the variation of the renormalization/factorization scale by a factor of $2$ from the central scale choice defined in eq. (\ref{scale}) (see below).}
\label{xsec-tHch}
\end{figure}

For the case of a heavy charged Higgs  boson, it is appropriate to 
describe $H^\pm$  production  in terms of the $bg\to tH^- $ + c.c. process \cite{Akeroyd:2016ymd}, so that the cross section is 
controlled by the ${H^+\bar{t} b}$ coupling. In fact, the decay process $H^\pm\to tb$ sees the intervention of the same coupling, though the 
corresponding dependence of the associated BR is somewhat washed away by the fact that this $H^\pm$ decay mode is the dominant one.  

The aforementioned vertex can be rewritten 
in a more convenient form using eq. (\ref{Yukawa-1}),
\begin{eqnarray}
 g_{\bar{t} bH^+} = i \left(C_L P_L + C_R P_R \right), 
 \label{Htb-coupling}
\end{eqnarray}
where $C_L = \frac{1}{\sqrt{2} v} m_t \kappa_u^A$, 
$C_R = \frac{1}{\sqrt{2} v} m_b \kappa_d^A$, wherein $\kappa_{u,d}^A$ are given in Tab. \ref{Yukawa-2} for the four different types of  2HDM. The nature of the chiral structure of the $H^+ \bar t b$ coupling is strongly dependent on $\tan\beta$. Firstly, in the 2HDM-I (and 2HDM-X), both the R- and L-handed components are proportional to $1/\tan\beta$ and hence, given that $C_L \propto m_t$ while $C_R \propto m_b$, the $H^+\bar t b$ coupling is dominated by the L-handed component. In contrast, in the 2HDM-II (and 2HDM-Y), the L- and R-handed components of  the
$H^+\bar t b$ coupling behave differently depending on $\tan\beta$:
(i)
for $\tan^2\beta < m_t/m_b$($\tan^2\beta>m_t/m_b$) the coupling is dominated by the L(R)-handed component; 
(ii) for $\tan^2\beta\approx m_t/m_b$, the coupling is purely scalar 
 with no pseudoscalar component. 


The dependence of the chiral structure of the $H^+\bar  tb$ vertex and the   $H^\pm$ production cross section 
as a 
 function of $\tan\beta$ are shown in the left and right panels, respectively, of Fig. \ref{xsec-tHch}\footnote{The error bands in the cross section show the effect of the theoretical uncertainties due to factorization/renormalization scale variations. These uncertainties are of order $\pm 14\%$ for $m_{H^\pm} = 200$  GeV and decrease to about $\pm 9.7\%$ for $m_{H^\pm} = 500$ GeV.}. 
 We can see that the cross section in the 2HDM-Y falls to a dip around 
 $\tan^2\beta\approx m_t/m_b$  and then increases for large $\tan\beta$. 
 In the 2HDM-I, however, it decreases always for increasing $\tan\beta$. As for the $H^+\bar tb$ vertex,  we notice first that $C_L$ is decreasing as a function of $\tan\beta$  in both types of 2HDM (the corresponding lines in fact overlap). However, 
 in the 2HDM-Y, $C_R$ can be about three orders of magnitude larger than in the 2HDM-I for large $\tan\beta$. Here, the importance of the L-handed part in the 2HDM-I through its effect on the $H^\pm$ production cross section will induce us to choose $\tan\beta$ as small as possible. Conversely, in the 2HDM-Y, for large values of $\tan\beta$, the $H^\pm$ production cross section can be three times higher than for low $\tan\beta \simeq 1$--$2$ and the R-handed part of the  $H^+\bar t b$ coupling can be about two orders of magnitudes higher than the L-handed part. 
 Therefore, here, we will choose a large value of $\tan\beta$.
 Such large $\tan\beta$ is not in conflict with theoretical constraints
 since one can always tune the Higgs masses and $\mu_{12}^2$ parameters for such purpose.
 
 For definiteness, in what follows,  we will choose our benchmark scenarios with $\tan\beta=1$ for the 2HDM-I and $\tan\beta=50$ 
 for the 2HDM-Y, which give
\begin{itemize}
    \item $(C_L,C_R) = (0.94, -0.025)$ for 2HDM-I.
    \item $(C_L,C_R) = (0.019, 1.3)$ for 2HDM-Y.
\end{itemize}
We then study three charged Higgs boson masses: $m_{H^\pm} = 300, 400$ and $500$ GeV for 2HDM-I plus  $m_{H^\pm} = 500, 600$ and $700$ GeV for 2HDM-Y.
\section{Phenomenological setup} 
\label{sec:setup}
We discuss here our analysis setup.\\

\paragraph*{\textbf{Observables.}} As discussed in section \ref{sec:2HDM}, $\tan\beta$ and $m_{H^\pm}$ are the main parameters controlling the production of a charged Higgs boson in association with a top quark and the spin properties of the top quark in this process. The former affects directly the chiral structure of the produced top quark while the latter controls its kinematics. In this section, we discuss the observables that we have used in this study to quantify the sensitivity of this production process to
top polarization effects, as function of $\tan\beta$ and $m_{H^\pm}$, 
in turn enabling one to pursue two goals: (i) to
distinguish the charged Higgs boson 
 signal from the SM background, (ii)  
 disentangle  different Yukawa types of the 2HDM from each other. 
 
 As previously explained, the properties of top (anti)quarks emerging at the production level are transmitted to its decays product, i.e., such a quark state decays before hadronizing. Hence, one can study the 
 differential distribution 
 in $\cos\theta_\ell^a$ of the emerging lepton, 
\begin{equation}
   \frac{1}{\sigma} \frac{\text{d}\sigma}{\text{d}\cos\theta_\ell^a} = 
   \frac{1}{2}\bigg(1 + \alpha_{\ell^\pm} P_{t,\bar{t}} \cos\theta_{\ell}^a\bigg),
 \label{theta}
 \end{equation}
wherein $\alpha_{\ell^\pm}$ is the so-called spin analyzing power of the  charged lepton and $\theta_{\ell^a} = \measuredangle (\hat{\ell}^\pm, \hat{S}_a)$, with $\hat{\ell}^\pm$ being the direction of flight of the charged lepton in the top quark rest frame and $\hat{S}_a$ the spin quantization axis in the basis $a$. There are three such bases relevant for top quark physics at the LHC (see, e.g., \cite{Bernreuther:2013aga} for details about the sensitivity and \cite{Aaboud:2016bit} for a corresponding measurement in $t\bar{t}$ production): the helicity basis, the transverse basis and the $r$-basis. In the helicity basis, $\hat{S}_a$ is the direction of motion of the top quark in the so-called ($t\bar{t})$ Zero Momentum Frame (ZMF). This basis will be denoted by the superscript $a=k$. In the transverse basis, the spin quantization axis is defined to be orthogonal to the production plane spanned by the (anti)top quark and the beam axis. Finally, the spin quantization axis in the $r$-basis is defined to be orthogonal to those in the helicity and transverse bases. The transverse and $r$-bases, although very useful for anomalous 
chromo-magnetic and chromo-electric top quark coupling studies (see, e.g., \cite{Bernreuther:2013aga}), are not a very sensitive probe in
the process we are considering here. Therefore, we limit this study to the helicity basis.

It was found that, in addition to angular observables,  energy distributions of the decay products of the (anti)top quark in the laboratory frame and their combination are excellent probes of top quark polarization. This is due to the fact that they acquire a dependence on $\cos\theta_X$ (where $X$ is a label of the decay product of the top quark) from boost factors, i.e., factors which bring the decay product from the top quark rest frame to the ($pp$) Center-of-Mass (CM) frame. Furthermore, they are sensitive to both the production and decay stages of the top quark. The first 
two observables are related to the $b$-jet energy in the laboratory frame 
\begin{equation}
u = \frac{E_\ell}{E_\ell+E_b}, \qquad
z = \frac{E_b}{E_t},
\end{equation}
where $E_\ell$, $E_b$ and $E_t$ are the energies of the charged lepton, 
$b$-jet and  top quark in the CM frame.
These observables were studied by the authors of \cite{Godbole:2011vw} to
distinguish $tW^\pm $ from $tH^\pm $ production. Then, they were supplemented by
 another observable for the purpose of  probing top quark $W^+\bar tb$ 
anomalous couplings in general \cite{Prasath:2014mfa} and specifically in single top production through the $t$-channel \cite{Jueid:2018wnj}. This is constructed as 
\begin{eqnarray}
 x_\ell = \frac{2 E_\ell}{m_t}, 
 \end{eqnarray}
where $E_\ell$ is the energy of the charged
lepton in the  CM frame and $m_t$ is the (pole) mass of the top quark.

\paragraph*{\textbf{Monte Carlo (MC) event generation.}} Events are generated here  at Leading Order (LO)  
using \textsc{Madgraph5\_aMC@NLO} \cite{Alwall:2011uj}\footnote{Note that Next-to-LO (NLO) QCD effects are not particularly relevant for studying the features of the top-bottom-charged Higgs vertex, so we have ignored these here.}
with the \textsc{NNPDF3.0} Parton Distribution Function (PDF) set using  $\alpha_s(M_Z^2)=0.118$ \cite{Ball:2014uwa}
for both the signal and  SM background.
The generated events were decayed with \textsc{MadSpin} \cite{Artoisenet:2012st} to keep full spin
correlations between the two produced top (anti)quarks and between each top (anti)quark
and its decay products. For the EW parameters, we have used the $G_\mu$-scheme in which 
the input parameters are $G_F, \alpha_\textrm{em}$ and $m_{Z^0}$. From these parameters, the 
values of $m_{W^\pm}$ and $\sin^2\theta_W$ are computed. To this end, we have used the following numerical inputs:  $G_F = 1.16639 \times 10^{-5} ~\text{GeV}^{-2}$, $\alpha_\textrm{em}^{-1}(0)=137$ and
$m_{Z^0}=91.188 \text{ GeV}$. For the fermion (pole) masses, we have  $m_t = 172.5$ GeV and $m_b = 4.75$ GeV.

The decayed events are then 
passed to \textsc{Pythia8} \cite{Sjostrand:2014zea} to add parton  shower and hadronization. We have further 
used the package \textsc{Rivet} \cite{Buckley:2010ar} for a particle  level 
analysis. Jets are clustered using \textsc{FastJet} \cite{Cacciari:2011ma}. 

Finally, we have used a dynamical choice for the renormalization/factorization scale, 
\begin{eqnarray}
 \mu_R=\mu_F = \frac{1}{2} \sum_{i=1}^N \sqrt{m_i^2 + p_{T,i}^2},
 \label{scale}
\end{eqnarray}
which is nothing but  the transverse mass of the final state divided by $2$. \\

\paragraph*{\textbf{Top quark reconstruction.}} In our analysis, the  reconstruction of the top (anti)quark was performed based on
the \textsc{PseudoTop} definition \cite{Collaboration:2267573} used widely by the ATLAS and CMS collaborations. Such a method was known to be resilient against Initial State Radiation (ISR)  that contaminates  the event sample. Our analysis used the \textsc{Rivet} implementation of the CMS measurement of the $t\bar{t}$ differential cross section at $\sqrt{s}=8$ TeV \cite{Khachatryan:2015oqa} with a slight modification on the lepton  and jet $p_T$ thresholds (see below). Further, in this study, we required a dressed lepton with cone radius $R_{\ell\gamma} = 0.1$. Two approaches are generally used in the reconstruction: the invariant mass based approach and $\Delta R$ based approach. We used the former, where all  jets (including $b$-jets), leptons (not from $\tau$ decays though), photons (to dress the selected leptons if they satisfy $\Delta R_{\ell\gamma} = \sqrt{(\eta_\ell - \eta_\gamma)^2 + (\phi_\ell - \phi_\gamma)^2} < R_{\ell\gamma}$) and missing energy are included. The  quantity 
\begin{eqnarray}
K^2 = \left(M_{\Tilde{t}_\ell} - m_t \right)^2 + \left( M_{j_1 j_2} - m_{W^\pm} \right)^2 + \left( M_{\Tilde{t}_h} - m_t \right)^2
\label{top-reconstruction}
\end{eqnarray}
is thus minimized to select the hadronic and leptonic top (anti)quarks. In eq. (\ref{top-reconstruction}), $\Tilde{t}_\ell$($\Tilde{t}_h$)
is the momentum of the (anti)top constructed in the leptonic(hadronic) decays from the $W^\pm$ boson decays and jet candidates. The minimization procedure adopted in this study can resolve the ambiguity of selecting jets and leptons correctly, e.g., whether a $b$-jet is coming from a semi-leptonic or  hadronic top pair decay. 

\section{Results}
\label{sec:results}
Before we illustrate our results, we first discuss our event selection and signal significance.\\

\paragraph*{\textbf{Event selection.}} Events are selected if they contain exactly one isolated charged lepton (electron or muon), at least $4$ jets where at least $2$ of them are $b$-tagged and missing transverse energy (which corresponds to the SM neutrino from $W^\pm$ boson decays). Electrons and muons coming from tau decays are not selected as events with one $\tau$ or more are rejected. We require the presence of one electron(muon) with $p_T > 30$ GeV($p_T > 27$ GeV) and $|\eta|< 2.5$($|\eta| < 2.4$). The missing transverse energy  is required to satisfy $E_T^\textrm{miss} > 20$ GeV. Jets are clustered using the anti-$k_T$ algorithm \cite{Cacciari:2008gp} with jet radius $\Delta R=0.5$ (as in the CMS analysis of \cite{Khachatryan:2015qxa}). We first require $p_T > 30$ GeV and $|\eta| < 2.4$ for all the jets in the events. We then refine our selection criteria by vetoing events which do not have a leading jet with $p_T > 50$ GeV. Finally, we select events that contain at least $5$ jets where at least 3 of them are $b$-tagged. This set of cuts will be denoted by \textsc{Cuts1}. Two more additional cuts on the $H_T$ quantity defined by 
\begin{eqnarray}
H_T = \sum_{i \in \textrm{jets}} p_T^i
\end{eqnarray}
were further imposed. The rationale for this is as follows. Background processes have a low peak in the $H_T$ distribution compared to the signals, especially for heavy charged Higgs bosons. Hence, imposing cuts on $H_T$ will improve significantly (beyond the basic selections) the signal-to-background ratio. We impose $H_T > 500$ GeV denoted by \textsc{Cuts2} and $H_T > 1000$ GeV denoted by \textsc{Cuts3}. We adopt  these last two cuts as representative of those that more refined selections may adopt in the actual $H^\pm$ signal search, still allowing for   the latter being enhanced with respect to the background  yet without biasing our MC data samples in the direction of removing completely the SM background (and its 
spin dependent features), as we want to benchmark the $H^\pm$ signal against it.

\noindent
\paragraph*{\textbf{Signal significance.}} For the production of a charged Higgs boson in association with a top quark followed by the $H^\pm \to t b$ decay, where one top decays hadronically and the other one leptonically, there are many background contributions. The most important ones are the exclusive production of a top (anti)quark pair in association with a $b$-quark (i.e., $t\bar t b$ + c.c.) and $t\bar{t}$ inclusive production. The first one is completely irreducible while the second one is partially reducible since the production of additional $b$-quarks is possible from the parton shower, notably in $g\to b\bar{b}$ splitting, but it is not a leading effect. There are further background processes possible, such as single top, di-boson and $W$ + jet production, but these are generally negligible compared to the two previous ones. 

Using  the described top (anti)quark reconstruction procedure on both heavy flavor states combined with  the requirements on jet activity and the cuts in $H_T$ will reduce substantially the background. To enable the possible observation of a signal, we compute its significance defined as \cite{Cowan:2010js}
\begin{eqnarray}
\mathcal{Z} = \sqrt{2 \bigg((N_s+N_b) \log\bigg(1+\frac{N_s}{N_b}\bigg) - N_s\bigg)},
\end{eqnarray}
where $N_s$($N_b$) is the number of signal(background) events 
after a given selection. We compute $\mathcal{Z}$ for $\mathcal{L} = 200$ and $1000$ fb$^{-1}$ of  integrated LHC luminosity. The obtained values are  displayed in Tab. \ref{table:significance}. We can see that already the basic selection (including the jet multiplicity requirement) enhances significantly the value of $\mathcal{Z}$. However, due to the small value of the cross sections for $m_{H^\pm} = 500$ GeV upwards, the cuts on $H_T$ are necessary to enhance significantly $\mathcal{Z}$ (at large luminosity). Indeed, we can see that, e.g., after \textsc{Cuts2}, $\mathcal{Z}$ increases by a factor of $\simeq 1.2$ for $m_{H^\pm} = 500$ GeV while it even decreases for lighter charged Higgs bosons. Before moving on, we should like to remind the reader here that these significances should not be regarded as a means of claiming discovery through our selection, rather as a means to indicate how to purify the signal so as to entertain the study of spin dependent observables that we illustrate now.

\begin{table}[!t]
\begin{center}
\begin{tabular}{| c || ccc | ccc |}
\cline{1-7}
 & \multicolumn{6}{|c|}{$\mathcal{L}=200(1000)$ fb$^{-1}$} \\
\cline{1-7}
& \multicolumn{3}{|c|}{\textsc{Type-I}} &
\multicolumn{3}{|c|}{\textsc{Type-Y}} \\ 
\cline{1-7}
\cline{1-7}
$m_{H^\pm}$ [GeV] &   $300$   & $400$ & $500$ &  $500$ & $600$ & $700$ \\ \hline 
Initial events &   $16.51~(36.92)$ &   $7.70~(17.22)$ & $3.87~(8.65)$ &     $7.33~(16.40)$  & $3.87~(8.67)$ &  $2.15~(4.79)$      \\ \hline
\textsc{Cuts1}   & $56.04~(124.79)$ & $30.45~(67.67)$ & $16.74~(37.52)$ &    $28.32~(63.78)$ &  $16.15~(36.45)$ &  $9.48~(21.09)$   \\ \hline
\textsc{Cuts2}   & $43.34~(96.42)$  &  $30.97~(69.01)$ &   $20.44~(46.02)$ &    $33.81~(76.10)$ &  $21.10~(47.50)$ &   $12.81~(28.50)$    \\ \hline
\textsc{Cuts3}   & $21.01~(45.68)$  &  $16.17~(35.18)$ & $12.21~(27.04)$ &  $19.63~(44.34)$ &  $15.18~(34.79)$ & $11.56~(26.15)$  \\ \hline
\hline
\end{tabular}
\end{center}
\caption{Signal significance of the two types of 2HDM that we study for $m_{H^\pm}=300, 400~ \text{and}~ 
500$ GeV in the 2HDM-I and $m_{H^\pm}=500, 600~ \text{and}~ 700$ GeV in the 2HDM-Y. The numbers outside(inside) the brackets refer to the case of $200~(1000)$ fb$^{-1}$ of total luminosity. }
\label{table:significance}
\end{table}

\begin{figure}[!h]
    \centering
    \includegraphics[width=0.48\linewidth]{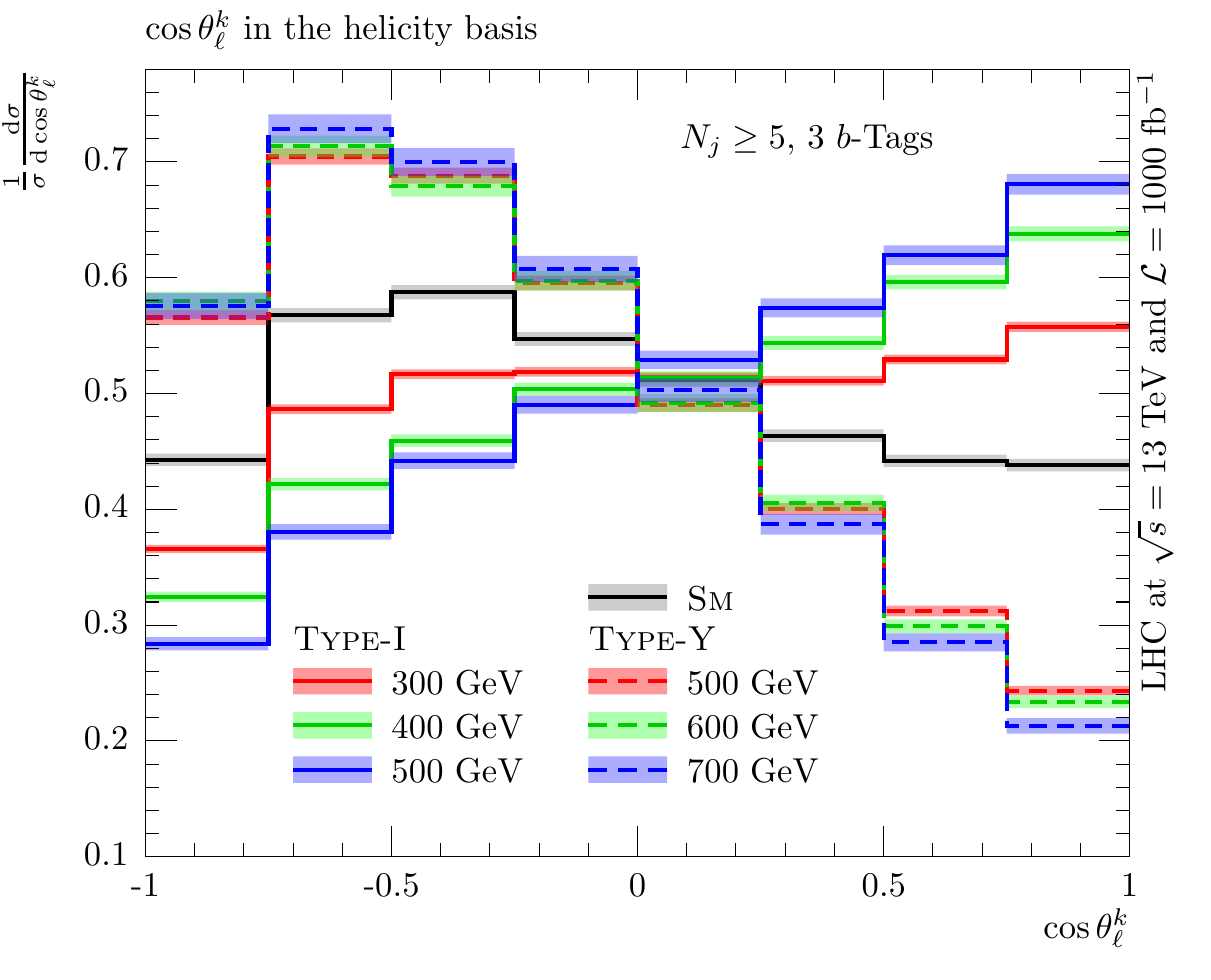}
    \hfill
  \includegraphics[width=0.48\linewidth]{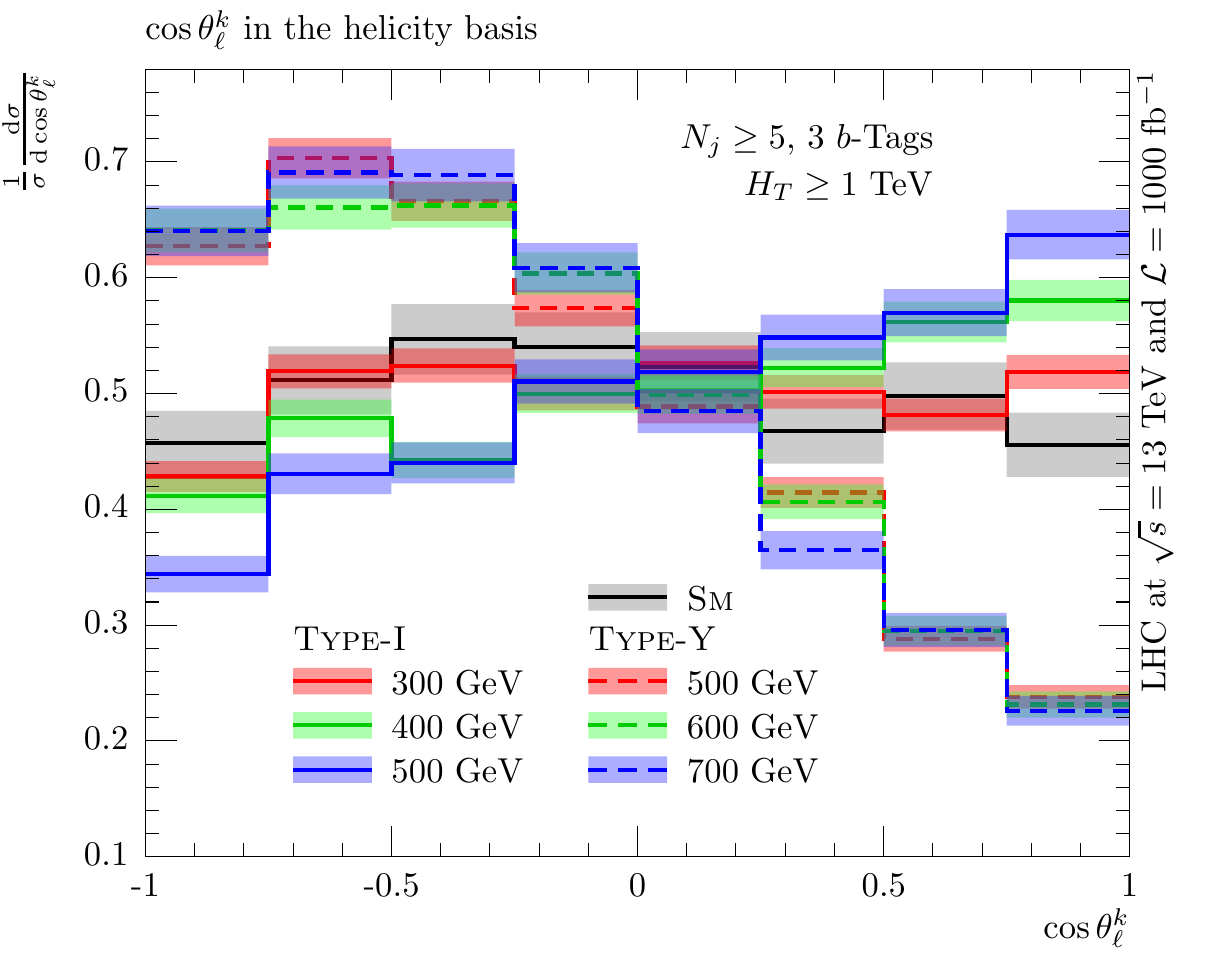}
    \caption{\emph{Left}: The $\cos\theta_\ell^k$ distributions for the SM (solid black), 2HDM-I (solid red, green and blue) and 2HDM-Y (dashed red, green and blue) after applying all the basic selections. \emph{Right}: the same distribution but for $H_T > 1$ TeV. Data are for $\sqrt s=13$ TeV and ${\cal L}=1000$ fb$^{-1}$.  The color scheme identify different $H^\pm$ masses while the shading represents the statistical and luminosity error.}
    \label{fig:fig2}
\end{figure}

\begin{figure}[tbp]
    \centering
    \includegraphics[width=0.48\linewidth]{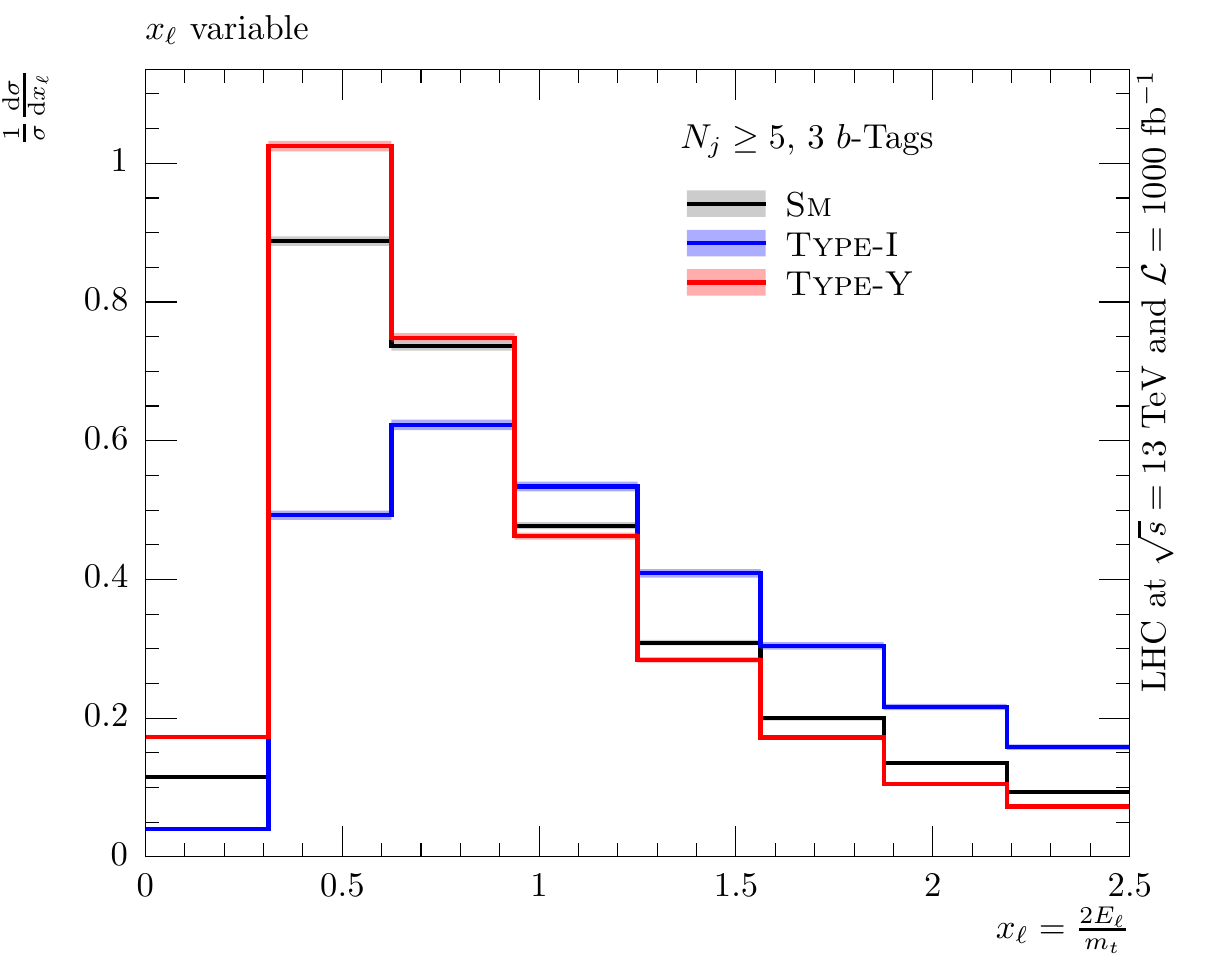}
    \hfill
  \includegraphics[width=0.48\linewidth]{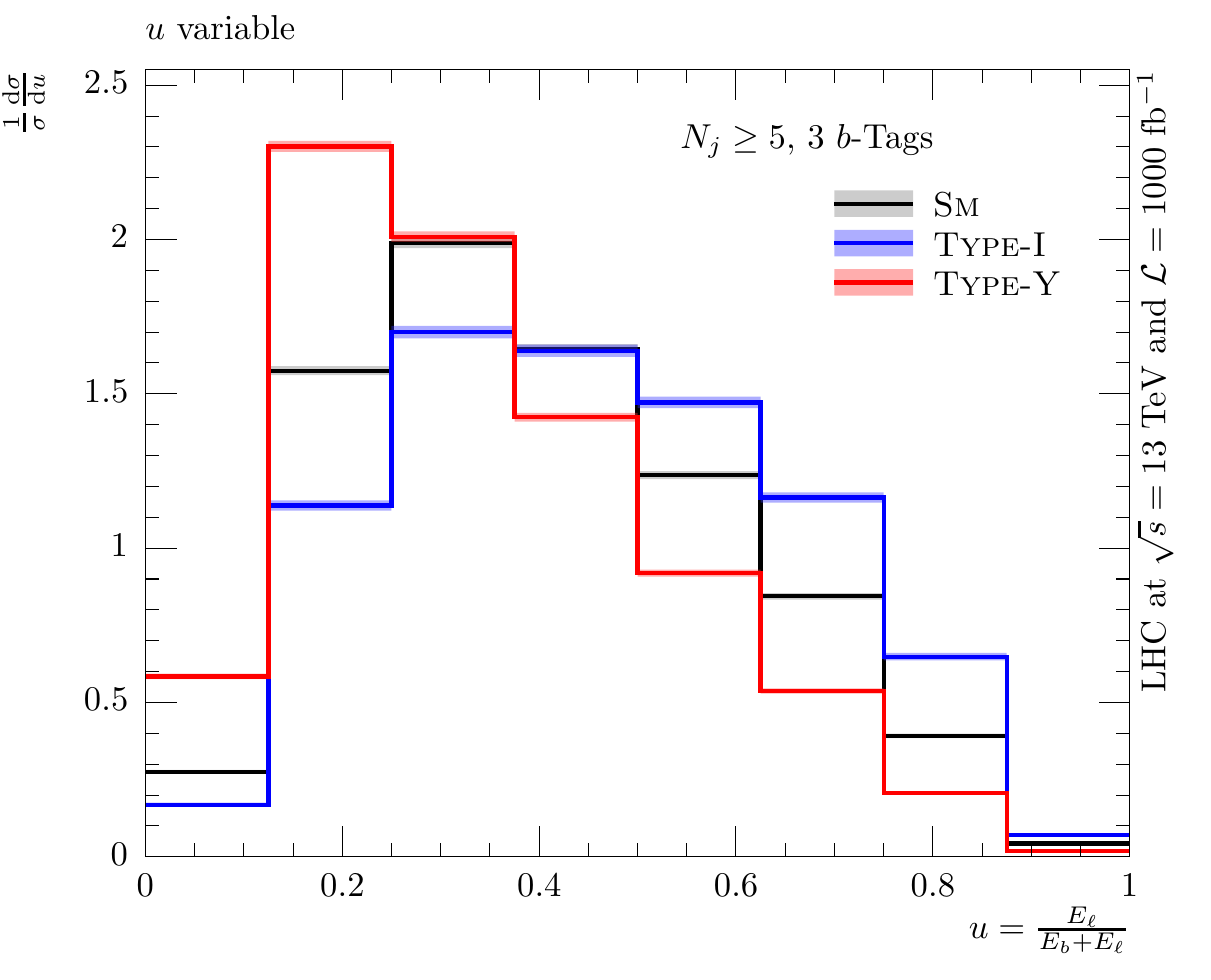}
    \caption{The $x_\ell$ (left panel) and $u$ (right panel) distributions for the SM (black), 2HDM-I (blue) and 2HDM-Y (red) applying the first set of cuts (denoted by \textsc{Cuts1}). Data are for $\sqrt s=13$ TeV and ${\cal L}=1000$ fb$^{-1}$.  The (almost invisible) shading represents the statistical and luminosity error. Here, $m_{H^\pm}=500$ GeV.}
    \label{fig:fig3}
\end{figure}

\noindent
\paragraph*{\textbf{Results.}} We present our results selecting three observables: $\cos\theta_\ell^k$ (i.e., in the helicity basis), $x_\ell$ and $u$. (We neglext showing the $z$ spectra, as they offer far less sensitivity in comparison.) The distributions are shown, after applying both the basic selection cuts (denoted by \textsc{Cuts1}) and the enhanced ones (e.g., \textsc{Cuts3}) in Figs.~\ref{fig:fig2} and \ref{fig:fig3}. 

In Fig. \ref{fig:fig2}, we show the $\cos\theta_\ell^k$ spectrum. The left panel shows the distribution after the \textsc{Cuts1} set. The label \textsc{Sm} shows the irreducible contribution to the background coming from $t\bar{t}b$ + c.c. process, which is the dominant one, though not the full SM noise (hence the different acronym). We can see clearly that the \textsc{Sm} curves exhibit almost no dependence on $\cos\theta_\ell^k$ except for regions of $\cos\theta_\ell \in [-0.75,0]$ where the EW contribution becomes important. This is unsurprising because it is otherwise dominated by a gluon intermediate state. Since this contribution is of vector nature, the polarization of the produced top quark is essentially  zero. The interesting observation is that the 2HDM-I (dominated in our benchmark by the L-handed component) and 2HDM-Y (dominated in our benchmark by the R-handed component) have opposite slopes and hence different polarization (with different sign). In the 2HDM-Y, the polarization of the top quark is negative while in the 2HDM-I it is positive. We remind the reader here  that the charged lepton is $100\%$ correlated with the parent top quark and hence the slope of the $\cos\theta_\ell^k$ distribution is directly proportional to the top quark polarization. Furthermore, the distribution has a drop for $\cos\theta_\ell^k \simeq -0.8$ due to the isolation cuts. Finally, we note that, after applying the cut $H_T > 1000$ GeV, one can see in the right panel of Fig.~\ref{fig:fig2} that the sensitivity decreases somewhat, especially for 2HDM-I, though it remains very noticeable. 

We should now also comment on the effect of changing the charged Higgs boson mass on the distributions of the angular observable. In the previous section, we have shown that $\tan\beta$ controls the chiral structure of the $H^+\bar{t} b$ coupling and hence directly its polarization. However, since  the latter is weighted by the cross section (via the helicity amplitudes entering the scattering matrix element and the phase space), we expect that it depends on the mass of the charged Higgs boson as well. We have checked this assumption for various values of the charged Higgs boson mass, i.e., $m_{H^\pm} = 300(500), 400(600)$ and $500(700)$ GeV for the 2HDM-I(2HDM-Y), and found that this is correct. First of all, for $m_{H^\pm} = 200$ GeV, there is almost no sensitivity to the different chiral structures of the 2HDM-I and 2HDM-Y, as both 
 models give approximately the same predictions and cannot be distinguished from the SM either. (This is why we do not show this case.) 
 However, starting from  $m_{H^\pm} = 300$ GeV, the studied observables start to exhibit the aforementioned  
 differences between the 2HDM-I, 2HDM-Y  and SM with a maximum sensitivity at large mass, e.g.,   
  $m_{H^\pm} = 500$ GeV for the 2HDM-I while for the 2HDM-Y the charged Higgs mass dependence is essentially negligible. Henceforth, then, we shall adopt $m_{H^\pm} = 500$ GeV as reference 
 charged Higgs mass value for our two BSM scenarios.
 
 In Fig.~\ref{fig:fig3}, we show the $x_\ell$ (left panel) and $u$ (right panel) spectra. The $x_\ell$ 
distribution shows no differences between the SM and 2HDM-Y and this is  due to cancellations among different terms proportional to the sign of the polarization, while for the 2HDM-I there are two discernible  features: position of the peak (which is slightly shifted respect to the SM) and behavior in the tail of the distribution. In contrast, $u$ is more sensitive and has a higher separation power than $x_\ell$ across the three theoretical setups, especially for the regions $0 < u < 0.25$ and $0.5 < u < 0.85$. In both these cases, the three scenarios are clearly distinguishable. Not shown here, we report that the effects of cuts on $H_T$ diminish somehow the sensitivity of both the $x_\ell$ and $u$ distribution to the underlying model assumption.

\section{Conclusion}
\label{sec:conclusion}

In summary, in the pursuit to establish a heavy charged Higgs boson signal at the LHC via its
$H^\pm\to tb$ decays, we have shown that spin dependent observables, both angles and
energy fractions of the (anti)top decay products, may improve the sensitivity of current analyses for  the purpose of both $H^\pm$ discovery and characterization, in fact, the former more than the latter, as they lead to forward-backward asymmetries that are clearly different in two types of 2HDM from the SM and from each other. We have proven this to be true for the case of two benchmark points over the parameter spaces of the 2HDM-I and 2HDM-Y, crucially differing in the chiral structure of the $H^+\bar t b$ vertex. Further, we have shown that such differences persist irrespectively of the $H^\pm$ mass value, so long that the latter is 300 GeV or more. Hence, just like the exploitation of spin effects has been fruitful over many years  for the case of $H^\pm\to \tau\nu$ decays (based on the pioneering work of Ref. \cite{Roy:1991sf}), we advocate a similar approach in the case of $H^\pm\to tb$ decays to now be taken.

We are confident that this will improve the LHC sensitivity to heavy charged bosons as we have reached  these conclusions by exploiting rather sophisticated phenomenological tools, adopting exact matrix element estimates, parton shower dynamics, hadronisation effects as well as jet definitions mimicking closely a typical experimental setup at the LHC, albeit we have not vigorously pursued a fully-fledged signal-to-background selection. We would expect this to be attempted now by ATLAS and CMS, possibly in conjunction with multi-variate analysis methods trained to learn the different spin dynamics affecting an $H^\pm$ induced signal and a $W^\pm$ dominated background.

Finally, we should close by remarking that, while we have illustrated all of the above by adopting two specific paradigms for BSM physics containing $H^\pm$ bosons, our approach can be exploited for a variety of other new physics frameworks containing such (pseudo)scalar charged states (e.g., {the Georgi-Machacek scenario, LR models, Supersymmetry, etc.}), so long that they induce chiral structures in the $H^+\bar tb$ vertex that are predominantly L- or R-handed.

\acknowledgments The authors would like to thank Marco A. Harrendorf for his collaboration in the early stage of this work. The work of AJ was supported by Shanghai Pujiang Program.
SM is supported in part through the NExT Institute and the STFC CG ST/L000296/1.  AA  and SM acknowledge funding via the H2020-MSCA-RISE-2014 grant no. 645722 (NonMinimalHiggs). 
AA acknowledges the Alexander von Humboldt Foundation and the Max Planck Institute for Physics, Munich, where part of this work has been done.

\bibliographystyle{JHEP}
\bibliography{biblio.bib}

\end{document}